\documentclass[18p,letter paper]{article}
\usepackage[utf8]{inputenc}
\usepackage{tensor}
\usepackage{amsmath}
\usepackage{amssymb}
\usepackage{amsfonts}
\usepackage{mathtools}
\usepackage{xfrac}
\usepackage{geometry}
\usepackage{graphicx}
\usepackage{epsfig}
\usepackage{subcaption}
\usepackage[numbers]{natbib}
\usepackage{color}
\date {September 2023}
\usepackage{multicol}
\usepackage{hyperref}
\usepackage{caption}
\hypersetup{
colorlinks=true,
breaklinks=true,
linkcolor=blue,
filecolor=green,
urlcolor=green,
citecolor=green,
bookmarksopen=true
}
\usepackage{authblk}

\begin{document}
\title{ \huge{Effective Hamiltonian of the k=+1 FRW model from the
K-Quantization method in LQC } \\ [10pt]}

    \author{ \large{Akram Hasseine}\footnote{email:akram.hasseine@doc.umc.edu.dz
}}
        
    \affil{Laboratoire de physique Mathématique et Subatomique,Faculty of science \\ Mentouri University,Constantine,Algeria}
    
\maketitle

\begin{abstract}
         In the effective theory of loop quantum cosmology LQC, the influence of the holonomy correction (with $\overline{\mu}$-scheme) on the homogeneous and the inhomogeneous cosmological models have been extensively studied in the case of flat space. In this paper,using the K-Quantization method and the $\overline{\mu}$-scheme,in the same framework of LQC,we construct the Hamiltonian constraint operator  of the  closed FRW model (k=+1), we show that the effective  semi-classical limit gives the same holonomy correction expression that was already used in the case of a flat space. We also derive the modified Friedman equation. The numerical investigation is performed to  plot the effective evolution  of the massless scalar field in term of the volume,which indicate a minimum volume at the bounce  and also a suitable infrared behavior.The condition that we apply to constrain the value of $\overline{\mu}$ in the semi-classical limit is consistent with the numerical results.
. 
\end{abstract}

\section{Introduction}

 The study of the universe evolution,or cosmology,is currently based on empirical data,probed by advanced satellites. It is still debated that recent data favor a closed universe. On one hand,recent results from the Planck collaboration \cite{aghanim2020planck} slightly suggest a closed universe, on the other hand it is concluded in \cite{di2020planck} that Planck data prefer positive curvature over $99\% C.L$ . Although there is no conclusive evidence for positive curvature,this remains at least a reasonable probability that is consistent with the data. Based on these recent results,our study focuses on the closed FRW model(k=+1);in the high-energy regime where the effects of quantum gravity and curvature  play an important role in the early universe. It is well known also that the universe with a closed model will end with the collapse when the density of matter reaches its minimum value.\\

 Studying the early universe  in high energy regime requires the use of a quantum theory of gravity .The application of Loop quantum gravity or LQG \cite{rovelli2015covariant,ashtekar2004background,thiemann2008modern}a candidate quantum theory of gravity, to the cosmological context; in particular; the quantization of the cosmological models is known as  Loop quantum  Cosmology or LQC\cite{bojowald2008loop,bojowald2003homogeneous,bojowald2003loop,ashtekar2006quantum}.

  Introducing a holonomy correction in the gravitational part of classical Hamiltonian constraint will then define an effective theory of Loop Quantum Cosmology; and this correction can be done to the homogeneous models \cite{bojowald2003homogeneous,bojowald2003loop,ashtekar2006quantum}  and it has led to a resolution of a big bung singularity ,as well as to the theory of inhomogeneous perturbation \cite{bojowald2008anomaly,cailleteau2012anomaly,bojowald2008loop,bojowald2006hamiltonian,bojowald2003loop,grain2010fully,date2004effective} to look at the influence of this corrections on the primordial power spectrum. In fact the observations of the fine imprint of this quantum effect by the future CMB missions allow us to test the model of LQC and also to understand the past of our universe \cite{grain2010observing}.  \\
 
In the case of a flat space (k=0) the holonomy correction is done by the replacement :  
 $\Bar{k} \to {\sin( \mu \gamma \bar k )}/{\mu \gamma} $  in the classical Hamiltonian Constraint, with 
 ${\mu}=\sqrt{{A_{min}}/{ \bar{p}}}  $    and  
 $ \Delta=A_{min}=2\pi\sqrt{3}\gamma l_{p}^{2} $. \\
 What about the holonomy correction in the closed model?.The detail study of the closed model using holonomies of the connections\cite{ashtekar2007loop} has led to a resolution of the big bung singularity and their study has also made it possible to find the equation of evolution which has a viable infrared behavior.

In this work we study  the effective Hamiltonian of the closed model. We will use  the improved dynamic ($\overline{\mu}$-scheme)\cite{vandersloot2007loop}  and the K-Quantization method \cite{bojowald2003homogeneous} which is a viable  mathematical strategy  because of the SU(2) gauge fixing. 
This method used the holonomy of the extrinsic curvature as a basic variable instead of the holonomy of connection used in\cite{ashtekar2007loop}.Using holonomies is interesting because it is compatible with the gauge invariant ground state but it leads to a non standard Hilbert space. This non standard  Hilbert space of LQC   is defined as the Cauchy completion of the space of almost periodic functions (cylindrical functions) with the Haar measure.What is the most important is that the spin network functions (periodic functions in this case) are all orthogonal to each other and constitute an orthonormal basis of this space ;which leads to the fact that what is considered as a well-defined operator in this Hilbert space is the holonomy of extrinsic curvature while the curvature operator  does not exist.\\

 Therefore, in order to construct the Hamiltonian constraint operator,we first rewrite the classical Hamiltonian constraint using the holonomies of the extrinsic curvature instead of the curvatures variables;then we perform the usual quantization as in standard quantum mechanics. After that we construct the  Hamiltonian constraint operator,we implement the semi-classical limit for it and derive the effective Hamiltonian. The interesting point is that the holonomy correction that we derived takes a simple expression similar to the one already used for the flat space.
\\ 
Moreover, and in order to perform the numerical calculations we must first constrain the  $\overline{\mu}$  value in the semi classical limit  using the minimum area $ \Delta=A_{min}=2\pi\sqrt{3}\gamma l_{p}^{2} $ deduced from the LQG  theory. At the end of the third section of this article we will comment this choice because in reality, the holonomies that we will use in the case of the closed metric to find the Hamiltonian constraint operator do not allow to realize a closed loop. We can then accomplish the numerical results,and we will obtain a constant critical density $ \rho_{crit}=0.82 \rho_{pl},
$ we will show that during the contraction of the universe with a massless scalar field the energy increases until it reaches its maximum value leading to a bounce, and we will also find that the universe in a closed metric will end with a collapse at minimum energy density which is the classical infrared limit.\\

This paper is organized as follow:
in  section (\ref{section}) we begin with the classical framework, in order to make a loop quantization with a canonical quantization scheme;we need first to derive the classical Hamiltonian constraint using symmetric reduced connection-triad variables of the closed metric,and then obtain the correct Friedman equation and the classical evolution trajectory (the evolution of the scalar field with respect to the volume) 
in section(\ref{sectionA}):we begin with a brief presentation of  the non standard Hilbert space of LQC, and then construct the Hamiltonian constraint operator in terms of the volume and the holonomy of the extrinsic curvature that are well-defined operators in the non-standard Hilbert space .In section(\ref{sectionB})using the semi- classical limit of Hamiltonian operator obtained in the previous section  we derive the effective Hamiltonian constraint and the effective equation of motion.finally,we can extract the holonomy correction and then derive the modified Friedmann equation . In section(\ref{sectionC}) we will accomplish the numerical results and we will discuss the phenomenological implication.
 \section{Classical framework } \label{section}
We begin with the classical framework that will form the basis of the loop quantization of the close model (k=+1) with canonical quantization scheme.The goal of this section is to consider the symmetric reduced connection-triad variables of the closed metric, and then construct the classical Hamiltonian constraint  in terms of this reduced symmetric variables,after that we show that the vanishing of this constraint gives back the correct Friedmann equation of the closed model.in order to make a comparison later with the effective evolution, we also deduce the classical evolution of the massless scalar field in terms of the volume.

\subsection{Symmetric reduced connection-triad variables
} 
In this subsection we introduce the symmetric reduced connection-triad variables of the closed metric witch are the reduced variables LQG.
Loop quantum gravity (L.Q.G) describes the gravitational fields as SU(2) non Abelian gauge  fields using background-independent method. In the canonical fields the so called Ashteckar variables are given by
\begin{equation}
     \{A^{i}_{a}(x),E^{b}_{j}(y) \}=\gamma\kappa\delta^{b}_{a}\delta^{i}_{j}\delta^{3}(x-y) 
\end{equation}
where $\kappa=8\pi G $ and $\gamma :The \text{ Barbero-Immirzi parameter}$. \\
   These variables are analogues of the vector potential and the electric field in electrodynamics.
 In the cosmological application (L.Q.C) for closed F.R.W metric,the homogeneous metric is given by:
      \begin{equation}\label{eq:einstein}
       dS^2=-\bar{N}(t)^2+\alpha(t)_{ij}\bar {\omega}^{i}_{a}\bar {\omega}^{j}_{b}dx^{a}dx^{b}
   \end{equation}
 where  $\alpha(t)_{ij}$ are the dynamical components of the metric. $\bar{N}(t)^2$ is known as the lapse that represents the re-scaling freedom of the time coordinate and $\overline  {\omega}^{i}_{a}$   are the basis of the left-invariant one-forms which satisfy
\begin{equation} \label{eq:einstein2}
    d\bar{\omega}^{i}=-\frac{1}{2}C^{i}_{jk} \bar{\omega}^{i}\wedge\bar{\omega}^{k}
\end{equation}
where $C^{i}_{jk} $ are the structure constants of the isometric 
group and thus characterise the Bianchi model.   
In particular for the closed model k=+1 the structure constants can be taken as \cite{ellis1969class}
\begin{equation}C^{i}_{jk}=2\varepsilon^{i}_{jk}\end{equation}
In addition to the left-invariant one forms, for what follows we will also need a basis of vector fields $\bar{e}^{a}_{i}$ which are also left invariant.
The left-invariant  vector field have commutators which provide a representation of the lie algebra  under consideration.

\begin{equation}\label{equation5}
\left[\bar{e}^{a}_{i},\bar{e}^{a}_{j}\right]=C^{k}_{ij}\bar{e}^{a}_{k}
\end{equation}
The left invariant vectors fields are also dual to   
$ \bar {\omega}^{i}_{a}$ and satisfies
\begin{equation} \label{eq:einstein3}
     \bar {\omega}^{i}_{a}\bar{e}^{a}_{j}=\delta^{i}_{j}
\end{equation}
In the ADM formalism,the general form of the closed Friedman-Robertson-Walker metric is given by
\begin{equation}
    g_{ab}=a^{2}\bar{q}_{ab}=a^{2}\bar\omega_{a}^{i}\bar\omega_{b}^{i}=a^{2}\{d\Dot{\phi}^{2}+sin\psi^{2}(d\theta^{2}+\theta^{2}d\phi^{2})\}
\end{equation}

The other component of the metric tensor satisfying 
\begin{equation}
   g_{oo}=-\bar{N}^{2} +g_{ab}\bar{N}^{ab}\quad \text {and} \quad \bar{N}^{ab}=0 \\
   \quad \text {with} \quad g_{0a}=q_{ab}N^{b}=0
\end{equation}
and the lapse function is given by
\begin{equation}
\bar{N}=\sqrt{|\bar{p}|}=a
\end{equation}
Using the homogeneous metric (\refeq{eq:einstein}),we can conclude that $\alpha(t)_{ij}=a^{2}(t)\delta_{ij}$ ,and $a(t)$ is the dynamical scale factor.\\
The starting point of loop quantum cosmology  is to reduce the Ashteckar variables \cite{bojowald2008loop} ; that is the su(2) valued connection $A^{i}_{a}$ which is  canonically conjugated to the triad $E^{a}_{i}$ of density weight one.

\begin{enumerate}
    \item The first canonical variable is the triad. In loop quantum cosmology; isotropic homogeneous triads have the following form  \cite{bojowald2008loop}
\begin{equation} \label{eq:einstein4}
{E}^{a}_{i}=\bar{p}\sqrt{\bar{q}}\bar{e}^{a}_{i}
\end{equation} 
 where $\bar{p}$ represents the dynamical component of the triad.
${E}^{a}_{i}$ encodes the spatial geometry in a specific fashion that is related to the spatial three metric through

\begin {equation}
    {E}^{a}_{i}{E}^{bi}=q^{ab}\det{q_{ab}}.
    \end{equation}
    \item The second canonical variable is the Ashteckar connection   
  \begin{equation}
     {A}^{i}_{a}={\Gamma}^{i}_{a}+\gamma{K}^{i}_{a}
 \end{equation}
 where ${K}^{i}_{a}$ is the extrinsic curvature and  ${\Gamma}^{i}_{a}$ is the spin connection given by
\begin{equation}
    \Gamma_{a}^{i}=-\frac{1}{2}\epsilon^{ijk}E_{j}^{b}\left(\partial_{a}E_{b}^{k}-\partial_{b}E_{a}^{k}+E_{k}^{c}E_{a}^{l}\partial_{c}E_{b}^{l}-E_{a}^{k}\frac{\partial_{b}(detE)}{detE}\right)
\end{equation}
The spin connection can also be written in function of the left invariant vectors and covector Fields 
     \begin{equation}
     \Gamma_{a}^{i}= -\epsilon^{ijk}e_{j}^{b}\left(\partial_{[a} e_{b]}^{k}+\frac{1}{2}e_{k}^{c}e_{a}^{l}\partial_{[a} e_{b]}^{l}\right)
    \end{equation}
    Using the last expression,direct calculations gives us
\begin{equation} \label{eq:einstein6}
    {\Gamma}^{i}_{a}=\bar{\omega}^{i}_{a}  
\end{equation}
which has a non vanishing value in the closed model.
And the extrinsic curvature ${K}^i_{a}$ is given by
\begin{equation} \label{eq:einstein7}
  {K}^{i}_{a}=\bar{k} \bar{\omega}^{i}_{a} \quad \text{with}\quad \bar{k}=sgn(\bar{p}).\frac{\dot{a}}{a}
\end{equation}
\end{enumerate}
\subsection{Classical Hamiltonian constraint}
With the symmetric reduced connection-triad variables, the next step is to show that the Hamiltonian formulation leads to the correct classical equation  of motion.The gravity part of the Hamiltonian constraint is given by:

\begin{equation} \label{eq:einstein 10}
    H_{G}=\frac{1}{2\kappa}\int_{\Sigma}d^{3}x\left(\bar{N}^{i}G_{i}+\bar{N}^{a}C_{a}+\bar{N}S\right)
\end{equation}
Which is the composition of three constraints\\
$G_{i}$ : is the Gauss constraint 
\begin{equation}
    G_{i}=D_{a}E^{a}_{i}=\partial_{a}{E}_{i}^{a}+\varepsilon_{ij}^{k}A^{j}_{a}E^{a}_{k} ,
\end{equation}
$C_{a}$ : is the diffeomorphism constraint
\begin{equation}
    C_{a}=E^{b}_{i}F^{i}_{ab}-(1-\gamma^{2})K^{i}_{a}G_{i} ,
\end{equation}
and the scalar constraint
\begin{equation}
    S=\frac{\bar{E}^{a}_{i}\bar{E}^{b}_{j}}{\sqrt{detE}}[\varepsilon^{ij}_{k}F^{k}_{ab}-2(1+\gamma^{2})K^{i}_{[a}K^{j}_{b]}]
\end{equation}
In the closed homogeneous and isotropic model k=+1 the Gauss constraint and the diffeomorphism constraint vanishe identically,leaving only the Hamiltonian (scalar) constraint.
The Hamiltonian constraint (20) without Immirzi parameter can also be written as\cite{bojowald2008anomaly}  
\begin{equation}
    S=\frac{\bar{E}^{a}_{i}\bar{E}^{b}_{j}}{\sqrt{detE}}\varepsilon^{ij}_{k}[2\partial_{a}\bar{\Gamma}^{k}_{b}+\varepsilon^{k}_{mn}(\bar{\Gamma}^{m}_{a}\bar{\Gamma}^{n}_{b}-\bar{K}^{m}_{a}\bar{K}^{n}_{b})]
\end{equation}
with
\begin{equation} \label{eq:einstein8}
    \sqrt{detE}=\bar{p}^{3/2}\sqrt{\bar{q}}
\end{equation}
In order to calculate  the Hamiltonian constraint(21), we can split the Hamiltonian expression into two parts; the Euclidean and the Lorentzian parts 
\begin{equation}
    S=C_{E}+C_{L}
\end{equation}
with \begin{equation}
    C_{E}=2\frac{\bar{E}^{a}_{i}\bar{E}^{b}_{j}}{\sqrt{detE}}\varepsilon^{ij}_{k}\partial_{a}\bar{\Gamma}^{k}_{b}
\end{equation}
and
\begin{equation} \label{eq:einstein9}
  C_{L}=\frac{\bar{E}^{a}_{i}\bar{E}^{b}_{j}}{\sqrt{detE}}\varepsilon^{ij}_{k}\varepsilon^{k}_{mn}(\bar{\Gamma}^{m}_{a}\bar{\Gamma}^{n}_{b}-\bar{K}^{m}_{a}\bar{K}^{n}_{b}).
\end{equation}
The straightforward calculation gives for the euclidean part
\begin{eqnarray}
   C_{E}=2\varepsilon^{ijk}\sqrt{|\bar{p}|}\sqrt{\bar{q}}\bar{e}^{a}_{i}\bar{e}^{b}_{j}\partial_{a}{\bar{\omega}^{k}_{b}}\nonumber&
    =&\varepsilon^{ijk}\sqrt{|\bar{p}|}\sqrt{\bar{q}}\bar{e}^{a}_{i}\bar{e}^{b}_{j}[\partial_{a}\bar{\omega}^{k}_{b}-\partial_{b}\bar{\omega}^{k}_{a}]\nonumber\\&=&
    \varepsilon^{ijk}\varepsilon^{kmn}\sqrt{|\bar{p}|}\sqrt{\bar{q}}\bar{e}^{a}_{i}\bar{e}^{b}_{j}[\bar{\omega}^{m}_{b}\bar{\omega}^{n}_{a}-\bar{\omega}^{m}_{a}\bar{\omega}^{n}_{b}]\nonumber\\&=&-12\sqrt{|\bar{p}|}\sqrt{\bar{q}}
\end{eqnarray}
where, after we have changed the spatial indices,in the first line , in the second line we have used the useful formula for the closed model
\begin{equation}
    \partial_{[a}\omega^{i}_{b]}=-\frac{1}{2}C^{i}_{jk}\omega^{j}_{[a}\omega^{k}_{b]}
\end{equation}
which can directly be  derived from the equation (\refeq{eq:einstein2}). In the third line of equation(26) we have used the orthogonality property given by the equation (\refeq{eq:einstein3}).\\
For the Lorentzian part,the direct calculation gives the expression
\begin{equation} 
    C_{L}=6(1-\bar{k}^{2})\sqrt{|\bar{p}|}\sqrt{\bar{q}}
\end{equation}
where we have used the equations (\refeq{eq:einstein4}),(\refeq{eq:einstein6}),(\refeq{eq:einstein7}),(\refeq{eq:einstein8})in the expression of the Lorentzian part(\refeq{eq:einstein9}).
Now combining the Euclidiean and Lorentzian parts we find for the gravitational part of the scalar constraint
\begin{equation}
    S=C_{E}+C_{L}=-6\sqrt{\bar{q}}\sqrt{|\bar{p}|}(1+\bar{k}^{2})
\end{equation}
Substituting the gravitational part of the the scalar constraint (29) into the Hamiltonian (\refeq{eq:einstein 10}) we get 
\begin{equation}
    H^{grv}=\frac{1}{2\kappa}\int_{\Sigma}d^{3}x\bar{N}S=-3V_{0}
    \frac{\bar{N}\sqrt{|\bar{p}|}}{\kappa}(\bar{k}^{2}+1)
\end{equation}
In the following we use the hamiltonian density
\begin{equation} 
    C_{gra}=\frac{H^{grv}}{\bar{N}}=\frac{1}{2\kappa \bar{N}}\int_{\Sigma}d^{3}x\bar{N}S=-3V_{0}
    \frac{\sqrt{|\bar{p}|}}{\kappa}(\bar{k}^{2}+1)
\end{equation}
Note  that the total Hamiltonian is composed of the the gravitationl and the matter part
\begin{equation} \label{eq(32)}
    H^{T}=H^{grav}+H^{m}=-3V_{0}\bar{N}
    \frac{\sqrt{\bar{p}}}{\kappa}(\bar{k}^{2}+1)+H^{m}
\end{equation}
Using 
\begin{equation}
    C^{T}=H^{T}=C^{grav}+C^{m}=\frac{H^{grav}}{\bar{N}}+\frac{H^{m}}{\bar{N}}
\end{equation}
and assuming that the matter Hamiltonian is only a function of $\bar{p}$ and not a function of $\bar{k}$ ( Note that we are only interested in the scalar or perfect fluids without fermionic fields).

\begin{equation}
\Dot{p} /2p=\Dot{a} /a=H \end{equation}

If we use the the Poisson bracket of the dynamical variables
\begin{equation} \label{reviewD}
    \{\bar{k},\bar{p}\}=\frac{\kappa
        }{3V_{0}}
\end{equation} 
we obtain
\begin{equation}
    \Dot{\bar{p}}= \{\bar{k},C^{T}\}=-\frac{\kappa
        }{3V_{0}}\frac{\partial C^{T}}{\partial\bar{k} }
\end{equation}
Using the fact that $\frac{\partial C^{m}}{\partial\bar{k} }=0$(for scalar or perfect fluids) with the formula (31) for $ C^{grav}$ we find

\begin{equation}
    \Dot{\bar{p}}=2\bar{k}\sqrt{\bar{p}} \quad \text{and} \qquad H^{2}=\frac{\bar{k}^{2}}{\bar{p}}
\end{equation}
Now we can use the vanishing of the total constraint to relate the right hand side of the gravitational density to the matter density using 
\begin{equation}
    C^{T}=C^{grav}+C^{m}=0 \Rightarrow	3V_{0}
    \frac{\sqrt{|\bar{p}|}}{\kappa}(\bar{k}^{2}+1)=C^{m}\end{equation}
  but 
  \begin{equation}
      C^{m}=V_{0}\bar{p}^{3/2}\rho_{m}
  \end{equation}
  recombining (37) and (38) and using the equation (39) we find
  \begin{equation}
      H^{2}=\frac{1}{3}\kappa\rho_{m}-\frac{1}{a^{2}}.
  \end{equation}
  Thus the reduced Hamiltonian gives back the correct Friedmann equation in the classical framowork.
   \subsection{Classical dynamic equation}
  We can now find the equation that shows the evolution of the volume in terms of the scalar field in the classical case.\\
     We have
    \begin{equation}
        \dot{v}=\{v,C^T\}_(\bar{k},\bar{p})=-\frac{\kappa}{3V_{0}}\frac{\partial{C^T}}{\partial{\bar{k}}}\frac{\partial{v}}{\partial{\bar{p}}}=3\bar{p}\bar{k}
        \end{equation}
   We note that $C^m$ does not depend on $\bar{k}$;we have only considered a scalar field but not a fermions field.
   we have
   \begin{equation} \label{42} 
     \rho_ {m}^{eff} =\frac{\bar{\Pi}_{\phi}^2}{2\bar{p}^3},
   \end{equation}
   that gives
   \begin{equation}
       C^{m}=\frac{V_{0}}{2\bar{p}^{3/2}}{\bar{\Pi}_{\phi}^2}
   \end{equation}
   Using the Poisson bracket for the scalar field we find the equation
     \begin{equation} \label{44}
    \dot{\Phi}=\{\Phi,C^T\}=\frac{1}{V_{0}}\frac{\partial C^T}{\partial{\bar{\Pi}_{\phi}}}=\frac{\bar{\Pi}_{\phi}}{\bar{p}^{3/2}}
   \end{equation}
   From the equations (41) and (44) we would derive the derivative of the volume with respect to the scalar field
   \begin{equation}
       \frac{dv}{d\phi}=3\bar{p}\bar{k}\frac{\bar{p}^{3/2}}{\bar{\Pi}_{\phi}}=\dot{v}\frac{\bar{p}^{3/2}}{\bar{\Pi}_{\phi}}=3Hv\frac{\bar{p}^{3/2}}{\bar{\Pi}_{\phi}}
   \end{equation}
       In the last equation we have used $\frac{\dot{v}}{v}=3H$.
   Using the vanishing of the total hamiltonian constraint 
   \begin{equation}
   C^{T}= C^{grav}+ C^{M}=0\implies 3V_{0}\frac{\sqrt{\bar{p}}}{\kappa}(\bar{k}^2+1)=V_{0}\bar{p}^{3/2}\rho_{m}^{eff}
   \end{equation}
   wich implies
      \begin{equation}
      \bar{k}=\left(\frac{1}{3}\kappa\bar{p}\rho_{m}^{eff}-1 \right)^{1/2}
     \end{equation}
   so that
   \begin{equation}
   H=\frac{\bar{k}}{\bar{p}^{1/2}}=\left(\frac{1}{3}\kappa\rho_{m}^{eff}-\frac{1}{a^2} \right)^{1/2}
   \end{equation}
   Substituting the last equation (48) in Eq.(45) we find
     \begin{equation}
      \frac{dv}{d\phi}=  \sqrt{12\pi G}\left(1-\frac{6v^{4/3}}{\kappa\bar{\Pi}_{\phi}^2} \right)^{1/2}v
     \end{equation}
       which can be integrated explicitly and gives  
       \begin{equation}
      \Phi(v)=\pm \frac{6}{\sqrt{12\pi G}}atanh\left(\sqrt{1-\frac{v^{4/3}}{\kappa\bar{\Pi}_{\phi}^2}}\right)+\Phi_0
     \end{equation}
       that  represents the classical trajectory( the evolution of the the scalar field ),where the positive solution indicates an expanding universe while the negative solution indicates a contracting universe. Note that this function is monotonic as it is shown in Figure 1,therefore it  can play the role of the emergent time.
        \begin{figure}[h]
    \centering
    \includegraphics [scale=0.5]{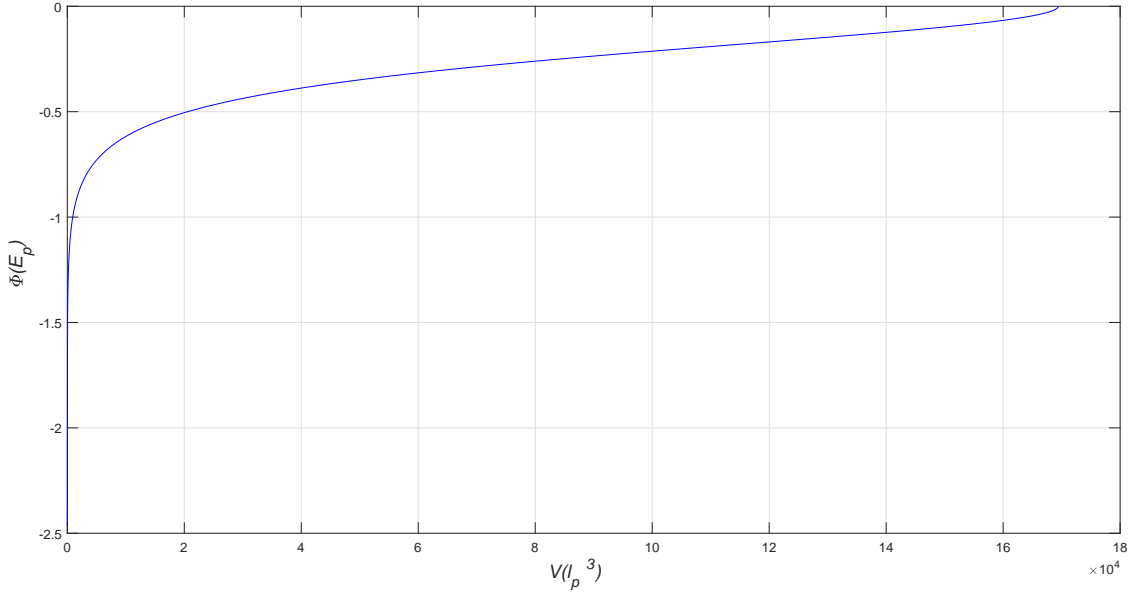}         \caption{monotonous evolution of the scalar field $\Phi$  with respect to the volume v .In this figure we have assumed $\bar{\Pi}_{\phi}=1500l_p$ for an expanding universe.}
    \label{fig:my_label}
    \end{figure}
    
   \section{Quantization} \label{sectionA}
    
   In this section we are interested to the loop quantization of our model. To achieve this in the canonical quantization scheme we follow the two steps:
   \begin{enumerate}
       \item Firstly, we need to choose a set of basic variables and finds a quantum representation of their algebra.
        \item Secondly, we need to construct an operator corresponding to the Hamiltonian constraint that is self-adjoint in the kinematic Hilbert space.
   \end{enumerate}
    
     \subsection{Kinematical Hilbert space}
  \subsubsection{Elementary variables }
  To construct the kinematical Hilbert space we must first  consider the elementary variables that will form the basis for quantization.
  From the full theory of loop quantum gravity we don't take the connection $A^{i}_{a}$ and  the triad $E^{a}_{i}$ as the  basic variables \footnote{The fact that we did not know to work with function of connection $ \Psi(A)$ that are,gauge and diffeomorpfism invariant in a controlled mathematical fashion, with a suitable (inner-product), and the difficulties of promoting the Hamiltonian constraint to an operator, in the full theory,the configuration variables are constructed from Holonomies and fluxes .Note that the non standard representation of L.Q.C followed naturally from the full theory by considering holonomies and  fluxes  violates a key assumption of Von Neumann's uniqueness theorem and gives the main difference between the Wheeler-Dewitt  quantisation and  that of LQC .}, rather, in the case of connection, one integrate $A^{i}_{a}$ along edges and then exponentiates the quantity leading to a holonomy \cite{ashtekar2003mathematical}. 
  
  The holonomy variables are then taken as the basic variable of the configuration space.
  The momentum variables are fluxes which are constructed by integrating the triad over a two surface.
  In cosmological setting fluxes are simply proportional to $\bar{p}$ which therefore forms an elementary variable.The construction of the k=+1 model presented here,follows closely that of k=-1 (open model)presented in\cite{vandersloot2007loop}.
  
  In the case of closed F.R.W space k=+1 the holonomy of connection takes a complicated form.However we can choose a K-Quantization method which is simpler.In the K-Quantization,the regulated Hamiltonian constraint is to be constructed using holonomies of the extrinsic curvature i.e holonomies of the connection minus the spin connection. First we begin with the parallel propagator along the oriented edge $l_i=l_0e^{a}_{i}\partial_a$ that is 
  
    \begin{eqnarray} \label{eq:einstein 11}
      h_{i}(\bar{A}^{j}_{a}-\bar{\Gamma}^{j}_{a})=h_{i}(\gamma\bar{K}^{j}_{a})=h_{i}(\gamma\bar{k}\bar{\omega}^{j}_{a})&=&\exp\int{ds}\gamma\bar{K}^{j}_{a}\bar{X^{a}_{i}}\tau_{j}\nonumber\\&=&\exp\int{ds}\gamma\bar{k}\tau_{i}\nonumber\\&=&\exp(l_{0}\gamma\bar{k}\tau_{i})
      \end{eqnarray}
      
            where $l_{0}=\bar{\mu}V_{0}^{1/3}$ in the ($\bar{\mu}$-scheme)
        is the length of the oriented edge with respect to the fiducial metric $\bar{q}_{ab}$. And $\tau_{i}=-\frac{i}{2}\sigma_{i}$ are the generators of the Lie algebra of SU(2) satisfying $[\tau_{i},\tau_{j}]=\epsilon^{k}_{ij}\tau_{k}$ .\\

In the case of the flat space the holonomies $ h_{i} h_{j}h_{i}^{-1}h_{j}^{-1}$ can be taken around a closed loop forming the unitary matrix that give the  measure of the flux of the field strength i.e. the curvature  across a surface bounded by the loop. Although in the case of k=+1 model the last Holonomies do not make a closed loop ,we can use it with a good approximation in large volume ($a\gg 1$)as we will show later at the end of the third section.
      
      The physical length of the edge depends on the physical geometry ; namely the value of the scale factor or the value of the triad $\Bar{p}$.
            \begin{equation}
       l_{phy}=l_{0}a=l_{0}{\bar{p}}^{1/2}=V_{0}^{1/3}\bar{\mu}\sqrt{\bar{p}}
     \end{equation}
      We assume that the parameter $\bar{\mu}$ appearing in the formula of the parallel propagator (\refeq{eq:einstein 11})  associated with extrinsic curvature is a function of $\bar{p}$ (improved dynamics):i.e $\bar{\mu} \propto 1/\sqrt{\bar{p}}$ and not a constant in contrast to the original literature of L.Q.C \cite{bojowald2003loop} where the $\mu_{0}-scheme$ with  ($\mu_{0} $) is constant; present the following issues
      
     1-The holonomies depend on the choice of the fiducial cell.
     
     2- A direct result of this in the k=0 model is that to the critical density implying a stronger quantum-gravity effect to the critical density as the universe expands in contrast with the observation.\\
     The $\bar{\mu}$ scheme\footnote {introduced firstly by 
 Kevin vandersloot\cite{vandersloot2007loop}} \cite{vandersloot2007loop} provides a solution to get out this issues ; i.e $\bar{\mu} \propto 1/\sqrt{\bar{p}}$, under this assumption, the critical density of the flat metric would be planckian and thus could only be consequential in the high energy regime of the early universe.\\
     For now let us assume that $\bar{\mu}$ is given by 
     \begin{equation} \label{eq53}
         \bar{\mu}=\frac{1}{V_{0}^{1/3}}\sqrt{\frac{\Delta}{\bar{p}}}
     \end{equation}
     where $\Delta$ is a constant to be fixed later. The formula ((\refeq{eq:einstein 11})) with $\bar{\mu}$ scheme is of course independent of the background metric $\bar{q}_{ab}$.
     \subsubsection{Preferred Kinematical Poisson subalgebra }
     Since we have used the holonomies on the classical configuration space,therefore the algebra generated by this holonomies is just the algebra generated by the almost periodic functions of the extrinsic curvature (the set of  cylindrical functions of the extrinsic curvature denoted by  $Cyl_{S}$) \cite{thiemann2008modern,ashtekar2003mathematical} 
       \begin{equation}
            g(\bar{k})=\sum_{j}\exp i\frac{(\mu_{j} \gamma V_{0}^{1/3}\bar{k})}{2},
 \quad \text where \quad {\xi_{j}}\in \mathbb C \quad \Bar{k}\in \mathbb R
        \end{equation}
     and the Poisson bracket between these elementary functions and the momentum p is:
     
     \begin{equation}
           \{g(\bar{k}),p\}=\frac{8\pi\gamma G}{6}\sum_{j}(i \mu_{j}{\xi_{j}}) \exp i\frac{(\mu_{j} \gamma V_{0}^{1/3}\bar{k})}{2}
        \end{equation}
        
     which forms the preferred Poisson subalgebra *-algebra $\beta$.
    \subsubsection{Representation of the corresponding Abstact *-Algebra $\beta$}
    The closure of the algebra *-algebra -of cylindrical functions denoted by $\overline{ Cyl_{S}}$ in the super-norm on R is an Abelian C*-algebra with respect to pointwise operations and a complex conjugation as involution. \\
        Now the Gel'fand theory garantees that \cite{thiemann2008modern,ashtekar2003mathematical} : 
    \begin{equation}
      \Delta(\overline{ Cyl_{S}}) \simeq \bar R \simeq \Delta(\textit{C}(\bar R)),      
    \end{equation}
     where $\Delta(\overline{ Cyl_{S}})$  is the spectrum of this algebra and is called the Bhor compactification of the real line
     $\bar R $ and the only continuous functions in $\bar R$  are the almost periodic functions \\
     The non standard Hilbert space $L^{2}(\bar{R}_{boh},d{\mu}^{H})$ is the Cauchy completion of the space $\overline{ Cyl_{S}}$ of almost periodic functions equipped with the Haar measure defined by\cite{bojowald2008loop} 
      \begin{equation}
        \int_{\bar R} f(c)d{\mu}^{H}=\lim_{T\to {\infty}} \frac{1}{2T}\int_{-T}^{+T}f(c)dc ,
         \end{equation}
         where in the right hand side the Lebesgue measure on R is used \\
     Using the last measure we can easily verify that the spin network functions which form an orthonormal basis in the Hilbert space $L^{2}(\bar{R}_{boh},d{\mu}^{H})$  are given by :
     
      \begin{equation}
        \langle \exp{i\frac{(\mu_{1} \gamma V_{0}^{1/3}\Tilde{\bar k})}{2}}| \exp{i\frac{(\mu_{2} \gamma V_{0}^{1/3}\Tilde{\bar k})}{2}} \rangle=\delta_{\mu_{1},\mu_{2}} \quad \text where \quad \Tilde{\bar k}\in \bar R \qquad
     \end{equation}
               or simply denoted by
         \begin{equation}
          \langle \mu_{1}   |\mu_{2}\rangle=\delta_{\mu_{1},\mu_{2}},
          \end{equation}
        where  $\delta_{\mu_{1},\mu_{2}}$ is the Kronecker delta, which demonstrates that all the eigenvectors are normalizable and orthogonal to each other and this is the characteristic of the non standard Hilbert space of LQC resulting from the use of holonomies as basic variables. 
          \\
          The Holonomy operator in the basis state is given by :
          \begin{equation} \label{E60}
           \widehat{\exp(i{\mu'}\Tilde{\bar{k}}/2)}|\mu\rangle=|\mu+\mu'\rangle.
     \end{equation}
    Thus,while the exponentials are well-defined operators, no operator for $\Bar{k}$ can be derived by taking a derivative of 
    $\exp(i{\mu}\bar{k}/2)$ by $\mu$  \footnote {for instance we can check that if $ \hat{\bar{p}}|\mu\rangle=\bar{p}|\mu\rangle $ and $\hat{{h}}_{\lambda}|\mu\rangle=|\mu+\lambda\rangle $ with $\hat{\bar{h}}_{\lambda}$  the holonomy opertor given by $\hat{h}_{\lambda}$ =$\hat{\exp{i}{\lambda}\bar{k}}$; since $|\mu+\lambda\rangle $is orthogonal to $|\mu+\lambda'\rangle $ for all $\lambda \neq \lambda'$ ;we conclude that $\hat{h}_{\lambda}|\mu\rangle$ is not continues with respect to$ \lambda$ then we can't find the derivative of  $\hat{h}_{\lambda}$ with respect to $\lambda$ witch would allow one to define the operator $\hat{\bar{k}}$; that is there is no operator corresponding to $\hat{\bar{k}}$}. Therefore  one way to understand the difference between the   Wheeler-Dewitt quantisation (WDW) and the LQC  quantization is to realize that while in the former theory the curvature operator is well defined in contrast to the latter theory,in the Hilbert space $L^{2}(\bar{R}_{boh},d{\mu}^{H})$ the eigenstates  of the $\hat{p}$ operator ,labeled by  $|\mu\rangle $ satisfy 
     
     \begin{equation}
              \hat{P}|\mu\rangle= -i \frac{\kappa\gamma\hbar}{3}\frac{\partial}{\partial \Tilde{\bar{k}} } |\mu\rangle= \frac{\kappa\gamma\hbar}{6}\mu|\mu\rangle.
                \end{equation}
     For the following it is useful to use the v representation; in this case the orthogonality condition (59) becomes 
      \begin{equation}
        \langle v_{1}|v_{2}\rangle=\delta_{v_{1},v_{2}}
     \end{equation}
   
     In this representation the volume operator is given by:
               \begin{equation} \label{E63}
              \hat{V}|v\rangle=\hat{P}^{3/2}|v\rangle=\left(\frac{\kappa\gamma\hbar}{6}\mu\right)^{3/2}|v\rangle={V}_v|v\rangle
          \end{equation}
       
    The parameter v runs over the entire real line, but the spectrum is discrete because of the orthogonality condition (62).The$|v\rangle$ basis are normalised and constitute a basis of kinematical Hilbert space $\mathcal{H}^{grav}_{kin}$.
     A general state in the kinematical Hilbert space can be expended as: 
     
     \begin{equation}
         |\Psi\rangle=\sum_{v}\Psi_{v}|v\rangle
     \end{equation}

           \subsubsection{ Hamiltonian constraint in term of Holonomy }

         In the classical framework, we have derived the Hamiltonain constraint(\refeq{eq(32)}). However we can not directly use this form of the constraint because it is expressed in terms of curvature  $\bar k $ itsef rather than Holonomies
     
     As mentioned in the last section,the fact that there is  no operator corresponding to $\hat{\bar{k}}$,the starting point for quantization which we will follow is to tack as the basis configuration  variables:the Holonomies  of the extrinsic curvature and the fluxes which are simply proportional to p. We conclude that $k^{2}$ term in the gravitational part of the Hamiltonian constraint (\refeq{eq(32)}) must be quantized, using the holonomy of the extrinsic curvature .\\
Following the result from the case (k=0), we can use the regularisation method already used by T.Thiemann  to define the required regulated Hamiltonian constraint operator.\\
Let us make the procedure more explicit.Considering the Taylor expansion of the holonomy (\refeq{eq:einstein 11}) along the oriented edge $l_i$ as $(\bar{\mu}\to 0$) 
\begin{eqnarray}
    \lim_{\bar{\mu}\to 0}h_{i}&=&\lim_{\bar{\mu}\to 0}\exp(\gamma\bar{k}\bar\mu V_{0}^{1/3}\tau_{i
        })\nonumber\\&=&1+\bar{k}\gamma\bar\mu V_{0}^{1/3}\tau_{i}+\frac{1}{2}(\bar{k}\gamma\bar\mu V_{0}^{1/3})^{2}\tau_{i}^{2}+O(\bar{\mu}^{3})
 \end{eqnarray}
from which we first find 
\begin{equation} 
    h_{k}\{h_{k}^{-1},{V}\}=-\frac{1}{2} sgn(\bar{p}) \kappa\gamma\bar\mu V_{0}^{1/3}\sqrt{|\bar p|} \tau_{k},
\end{equation}
after we have used the Poisson bracket of the classical dynamical variables(\refeq{reviewD}).\\
And we also find 
 \begin{equation}
     \frac{1}{\bar{\mu}^{2}}\sum_{ij}\varepsilon^{ijk}h_{i}h_{j}h_{i}^{-1}h_{j}^{-1}=2\gamma^{2}\bar{k}^{2}V_{0}^{2/3}\tau_{k}+O(\bar{\mu})
 \end{equation}
    Substituting  the equations (66) and(67) in the following expression
    \begin{equation} \label{reviewD2}
        C_{grav}=\frac{-2sgn(\bar{p})}{\kappa^{2}\gamma^{3}\mu^{3}}\sum_{ijk}\varepsilon^{ijk}tr[(h_{i}h_{j}h_{i}^{-1}h_{j}^{-1}+2\gamma^{2}\bar{\mu}^{2}V_{0}^{2/3}\tau_{i}\tau_{j})h_{k}\{h_{k}^{-1},\hat{V}\}],
    \end{equation}
   we can prove that when we tack the limit $(\bar{\mu}\to 0)$ we recover the classical Hamiltonian constraint from the last expression.\\
       \textit{Proof}:
     we can write (68) as:
    \begin{equation}
        C_{grav}=C_{grav}^{1}+C_{grav}^{2}
    \end{equation}
      \begin{enumerate}
          \item  For the first term,using the commutative property, we find
    \begin{eqnarray}
        C_{grav}^{1}&=&\frac{-2sgn(\bar{p})}{\kappa^{2}\gamma^{3}\bar{\mu}^{3}}\sum_{ijk}\varepsilon^{ijk}tr\left(h_{i}h_{j}h_{i}^{-1}h_{j}^{-1}\{h_{k}^{-1},\hat{V}\}\right)\\&=&\frac{-2sgn(\bar{p})}{\kappa^{2}\gamma^{3}\bar{\mu}^{3}}\sum_{k}\{\sum_{ij}\varepsilon^{ijk}tr\left[(h_{i}h_{j}h_{i}^{-1}h_{j}^{-1}\{h_{k}^{-1},\hat{V}\}\right] \},
    \end{eqnarray}
    
   now using the equations (66) and (67) we have (as $\bar{\mu}\to 0) $
    \begin{eqnarray}
     \lim_{\bar{\mu}\to 0} C_{grav}^{1}&=&  \frac{-2}{\kappa^{2}\gamma^{3}\bar{\mu}^{3}}tr \{\sum_{k}(2{\gamma}^{2}\bar{k}^{2}V_{0}^{2/3}\bar{\mu}^{2}\tau_{k})\times(-1/2\kappa\gamma\bar{\mu}\sqrt{\bar{p}}V_{o}^{1/3}\tau_{k})\}\nonumber\\&=&\frac{2\sqrt{\bar{p}}V_{0}\bar{k}^{2}}{\kappa}\sum_{k}tr(\tau_{k}^{2})=-\frac{3V_{0}\sqrt{\bar{p}}\bar{k}^{2}}{\kappa}
    \end{eqnarray}
      where we have used
                \begin{equation}\sum_{k}tr(\tau_{k}^{2})=-3/2\ .\nonumber\end{equation}
    
      \item For the second term in Eq.(68) we find  
    \begin{equation}
      C_{grav}^{2}=  \frac{-2sgn(\bar{p})}{\kappa^{2}\gamma^{3}\bar{\mu}^{3}}\sum_{ijk}\varepsilon^{ijk}\left[2\gamma^{2}\bar{\mu}^{2}V_{0}^{2/3}  tr(\tau_{i}\tau_{j}h_{k}\{h_{k}^{-1},\hat{V}\})\right],
    \end{equation}
   
     now using the expression (66) we have (as $\bar{\mu}\to {0}) $       \begin{eqnarray}
           \lim_{\bar{\mu}\to{0}} C_{grav}^{2}&=&\frac{-2}{\kappa^{2}\gamma^{3}\bar{\mu}^{3}}(2\gamma^{2}\bar{\mu}^{2} V_{0}^{2/3})\sum_{ijk}tr(\varepsilon^{ijk}\tau_{i}\tau_{j}\tau_{k})[\frac{-1}{2} \kappa\gamma\bar{\mu}V_{0}^{1/3}\sqrt{\bar{p}}]\nonumber\\&=&\frac{2V_{0}\sqrt{p}}{\kappa}\sum_{ijk}tr((\varepsilon^{ijk}\tau_{i}\tau_{j}\tau_{k})=\frac{-3V_{0}\sqrt{p}}{\kappa},
           \end{eqnarray}
        where we have used
             \begin{equation}
               \sum_{ijk}tr(\varepsilon^{ijk}\tau_{i}\tau_{j}\tau_{k})=-3/2\nonumber .
           \end{equation}
   \end{enumerate}
          Combining the results (72)and (74), then we get 
           \begin{equation}
               C^{grav}=\lim_{\bar{\mu}\to0}(C_{grav}^{1}+C_{grav}^{2})=-3V_{0}
    \frac{\sqrt{\bar{p}}}{\kappa}(\bar{k}^{2}+1)\nonumber.
           \end{equation} 
Thus, we find the classical expression of the Hamiltonian constraint after making the limit $(\bar{\mu}\to 0)$ of the new expression (\refeq{reviewD2}) which uses volume and holonmy as basic variables instead of the curvature variable. Note that this limit is correct despite the circulation using the composition of the holonomies $h_{i}h_{j}h_{i}^{-1}h_{j}^{-1}$ forms an unclosed loop because the left invariant vector fields $ \bar{e}^{a}_{i}$ defining the edges do not commute (see equation(\refeq{equation5})).We will see that in the  quantum theory the $\bar{\mu} $ cannot be shrunk to zero as it would require an operator for the extrinsic curvature which is not available,and the parameter $\bar{\mu} $ must be constrained using the result of the minimal surface in LQG.
                                                   
          \subsubsection{Quantum regulated Hamiltonian constraint }                      
           
      Now, we are ready to convert the classical expression into quantum-operators since both the holonomy and the volume operator are well defined in the kinematical Hilbert space.
      The expression (\refeq{reviewD2}) is promoted to an operator,and the classical Poisson bracket is replaced by the commutator bracket. as in standard quantum mechanic,
      \begin{equation}
       \{.,.\} \to -\frac{i}{\hbar}[.,.].
      \end{equation}
      Then we get 
      \begin{equation}  \label{E76}
        \hat{C}_{grav}=\frac{2isgn(\bar{p})}{\kappa^{2}\gamma^{3}\hat{\bar{\mu}}^{3}}\sum_{ijk}\varepsilon^{ijk}tr\{(\hat{h_{i}}\hat{h_{j}}\hat{h_{i}}^{-1}\hat {h_{j}}^{-1}+2\gamma^{2}\hat{\bar{\mu}}^{2}V_{0}^{2/3}\tau_{i}\tau_{j})\hat{h_{k}}[\hat{h_{k}}^{-1},\hat{V}]\}
    \end{equation}

    Now in order to make the semi classical limit of the quantum constraint operator in the next section, we don't tack the limit $(\bar{\mu}\to 0)$ due to the absence of the curvature operator as we have  mentioned in the last section and also  because in LQC the full diffeomorphism invariance is not available.  
    In the full theory LQG we shrink $(\bar{\mu}$ until the a closed loop spanned by the holonomies has the minimum area\cite{thiemann1998length} .In the case of closed space (k=+1) we cannot make a closed loop using holonomies $ h_{i} h_{j}h_{i}^{-1}h_{j}^{-1}$ but this result is almost identical to that of using holonnomies $ h_{i} h_{j}h_{i}^{-1}h_{j}^{-1}h_{[i,j]}^{-1}$ around a closed loop provided that we tack $\bar{\mu}=V_{0}^{-1/3}\sqrt{\Delta/ {\bar{p}}}$ with $\Delta=A_{min}$ and  ($\bar{p}\gg 1$), indeed  
         \begin{equation}
       h_{[i,j]}^{-1}=\exp(-2\epsilon_{ij}^k V_0^{2/3}\bar\mu^2 \gamma \Bar{k}\tau_k) \propto \exp(-\alpha_{ij}\frac{A_{min}}{a^2}\gamma \bar k) \approx 1, \quad \text where \quad  \bar{\mu}=V_{0}^{-1/3}\sqrt{\Delta/ {\bar{p}}},\quad \title \quad \text and\quad  a^2\gg 1.  	
        \end{equation}
    which means that $ h_{i} h_{j}h_{i}^{-1}h_{j}^{-1}h_{[i,j]}^{-1}\approx h_{i} h_{j}h_{i}^{-1}h_{j}^{-1} $ ,and using the same conditions indicated in the last equation we can also  guarantee that :
    \begin{equation}
          h_{i} h_{j}h_{i}^{-1}h_{j}^{-1}h_{[i,j]}^{-1}\approx h_{i} h_{j}h_{i}^{-1}h_{j}^{-1}\geq 1+\frac{1}{2}A_{min}\gamma^2 \bar{k}^2 
      \end{equation}
      This result also show that the area surrounded by the holonomies of the extrinsic curvature is bounded below by $A_{min}$,which is taken as evidence that we can constrain the 
 value of $\bar \mu $ in the expression (\refeq{eq53}) using $\Delta=A_{min}=2\pi\sqrt{3}\gamma l_{p}^{2}$ , and $\bar p=a^2\gg 1 $ to perform the K-Quantiztion of the closed model.

    \section{Effective semi-classical dynamics}\label{sectionB}
    
    We show that with the effective semi classical dynamics , we can obtain a modified Friedmann equation incorporating quantum gravity effects.
    Several methods have been used in the effective dynamic namely the technique of expectation values of kinematical coherent state\cite{ashtekar2003mathematical,willis2004low} and the WKB expansion of the solutions to the difference equation\cite{banerjee2005discreteness,date2004effective} and path integral considerations\cite{noui2005physical}.In this paper we use the most direct  \cite{vandersloot2005hamiltonian} used in the case of a flat space (k=0). 
    \subsection{ Effective Hamiltonian constraint}
    The  effective Hamiltonian constraint is given by \cite{noui2005physical,vandersloot2005hamiltonian}
    \begin{equation}
        H_{eff}=   \frac{\langle \bar{k}|\hat{H}|\bar{p}\rangle}{\langle\bar{k} |\bar{p}\rangle}.
    \end{equation}
    
    There are two remarks that must be taken into consideration for the construction of this effective Hamiltonian the first is that  $|\bar{k}\rangle$ is not an eigenstate of any $\hat{k}$ operator since none exists; rather it is an eigenstate of the holonomy operator $\hat{h_{i}}$.We conclude that $|\bar{k}\rangle$ and $|\bar{p}\rangle$ or ($|v\rangle$)are eignentates of the regulate operator in Eq.(\refeq{E76}),therefore we need only consider eigenvalues of the $\hat{V}$ and $\hat{h}$ operators in the regulated Hamiltonian constraint (\refeq{E76}) written in term of the classical dynamic variables $\bar{k}$ an  $\bar{p}$ to get an effective Hamiltonian constraint.The second remark is that in order to complete this construction we need to make two approximations, the first is due to the small extrinsic curvature and the second occurs for large volumes . \\
    
    Let us make the procedure more explicit.First we can write the operator (\refeq{E76})as
    
    \begin{equation}
        \hat{C}_{grav}=\hat{C}_{grav}^{1}+\hat{C}_{grav}^{2}
    \end{equation}
\begin{enumerate}    
    \item For the first part of the constraint 
    \begin{equation}
        \hat{C}_{grav}^{1}=\frac{2isgn(\bar{p})}{\kappa^{2}\gamma^{3}\hat{\bar{\mu}}^{3}}\sum_{ijk}\varepsilon^{ijk}tr\left(\hat{h_{i}}\hat{h_{j}}\hat{h_{i}}^{-1}\hat {h_{j}}^{-1}\hat{h_{k}}[\hat{h_{k}}^{-1},\hat{V}]\right),
    \end{equation}
      The similarity of the first term to the case k=0 (flat space) is more apparent.
     We can write the first term as 
        \begin{eqnarray} \label{82}
          \hat{C}_{grav}^{1}&=&\frac{2isgn(\bar{p})}{\kappa^{2}\gamma^{3}\hat{\bar{\mu}}^{3}}\sum_{ijk}\varepsilon^{ijk}tr\left(\hat{h_{i}}\hat{h_{j}}\hat{h_{i}}^{-1}\hat {h_{j}}^{-1}\hat{h_{k}}[\hat{h_{k}}^{-1},\hat{V}]\right)\nonumber\\&=& \frac{6isgn(\bar{p})}{\kappa^{2}\gamma^{3}\hat{\bar{\mu}}^{3}}tr\left((\hat{h_{1}}\hat{h_{2}}\hat{h_{1}}^{-1}\hat {h_{2}}^{-1}-\hat{h_{2}}\hat{h_{1}}\hat{h_{2}}^{-1}\hat {h_{1}}^{-1})\hat{h_{3}}[\hat{h_{3}}^{-1},\hat{V}]\right)
          \end{eqnarray}
        where we have used the fact that the operator is gauge invarian in the second line.
        Now we want to calculate the term $\hat{h}_{3}[\hat{h_{3}}^{-1},\hat{V}]$. For this object we choose the v representation; where the operator $\hat{V}$ acts on the $|v\rangle$of the $\mathcal{H}^{grav}_{kin}$ is given by Eq (\refeq{E63}).\\
           Now the expression $\hat{h}_{3}[\hat{h_{3}}^{-1},\hat{V}]$ in the v representation is given by 
          \begin{equation} \label{E82}
              \left(\hat{h}_{3}[\hat{h_{3}}^{-1},\hat{V}]\right)_{mn}|v\rangle=\left(\hat{V}\delta_{mn}-\hat{h_{3}}\hat{V}\hat{h}_{3}^{-1}\right)|v\rangle
          \end{equation}
          The explicit form of the matrix elements of the holonomies(In the $J=+\frac{1}{2})$ representation is given by
          \begin{equation}
              \left(h_{3}^{J=\frac{1}{2}}\right)_{mn}=e^{{im}{\bar{\mu}}\bar{k}\gamma V_{0}^{1/3}}\delta_{mn} ,
          \end{equation}
          where m and n are indices of matrix elements which vary from J=+1/2 to J=-1/2. Using the equation(\refeq{E60})the action of the operator $\hat{h}_{3}$ on the states $|v\rangle$ (states of $\mathcal{H}^{grav}_{kin}$) is simply given by:
        \begin{equation}
            \widehat{\exp{({im}{\bar{\mu}}\gamma V_{0}^{1/3}}\bar{k})}|v\rangle=|v+2m\bar{\mu}\rangle,
        \end{equation}
        Using the last expression and the equation(\refeq{E63}) we can now calculate the expression (\refeq{E82}) 
        \begin{eqnarray}
             \left(\hat{h}_{3}[\hat{h_{3}}^{-1},\hat{V}]\right)_{mn}|v\rangle\nonumber&=&\left(\hat{V}\delta_{mn}-\hat{h_{3}}\hat{V}\hat{h}_{3}^{-1}\right)|v\rangle\\\nonumber&=&V_{v}\delta_{mn}|v\rangle-\hat{h}_{3}\left( \hat{V}|v-2m\bar{\mu}\rangle\right)\\&=&\left(V_{v}\delta_{mn}-\delta_{mn}V_{v-2m\bar{\mu}}\right)|v\rangle  .
          \end{eqnarray}
          We  conclude that in the v representation and in J=+1/2 we have the matrices elements
          \begin{equation} \label{87}
               \left(\hat{h}_{3}[\hat{h_{3}}^{-1},\hat{V}]\right)_{mn}=\begin{pmatrix}V_{v}-V_{v+\bar{\mu}}&0\\0&V_{v}-V_{v-\bar{\mu}}\end{pmatrix}
          \end{equation}
            Now we use the second order approximation of the holonomy  to find the small curvature approximation in the curvature expression.
            
              \begin{equation}
              h_{i}=e^{\bar{\mu}\bar{k}\gamma V_{0}^{1/3}\tau_{i}}=1+\bar{\mu}\bar{k}\gamma V_{0}^{1/3}\tau_{i}+\frac{(\bar{\mu}\bar{k}\gamma V_{0}^{1/3})^{2}}{2}\tau_{i}^{2}+O(\mu^{3})
          \end{equation}
         Using the formula (87) and we consider the fact that in  this effective classical constraint we only consider eigenvalues of operators,then we simply remove the hates from the holonomies operators we find that
        \begin{equation} \label{89}
            h_{1} h_{2}h_{1}^{-1}h_{2}^{-1}-h_{2} h_{1}h_{2}^{-1}h_{1}^{-1}=2\left(\bar{\mu}\bar{k}\gamma V_{0}^{1/3}\right)^{2}\tau_{3}+\left(\bar{\mu}\bar{k}\gamma V_{0}^{1/3}\right)^{3}(\tau_{2}-\tau_{1})-2/3\left(\bar{\mu}\bar{k}\gamma V_{0}^{1/3}\right)^{4}\tau_{3}+O(\bar{\mu})
        \end{equation}
        Substituting the expressions (\refeq{87}) and(\refeq{89}) in the expression (\refeq{82}) we find:
        \begin{eqnarray}
             C_{grav}^1&=&\frac{6isgn(\bar{p})}{\kappa^{2}\gamma^{3}{\bar{\mu}}^{3}}\left[2\left(\bar{\mu}\bar{k}\gamma V_{0}^{1/3}\right)^{2}-2/3\left(\bar{\mu}\bar{k}\gamma V_{0}^{1/3}\right)^{4}\right]tr\left[\tau_{3}\begin{pmatrix}V_{v}-V_{v+\bar{\mu}}&0\\0&V_{v}-V_{v-\bar{\mu}}\end{pmatrix}\right]\\ \nonumber&=&\frac{-3}{\kappa\gamma^{2}}\left[\left(\bar{k}\gamma V_{0}^{1/3}\right)^{2}-1/3\bar{\mu}^2\left(\bar{k}\gamma V_{0}^{1/3}\right)^{4}\right]\times\frac{2}{8\pi \gamma l_{p}^2\mu}\left(V_{v+\bar{\mu}}-V_{v-\bar{\mu}}\right)
        \end{eqnarray}
        where we have used the fact that $ (\tau_{1})_{mn} \text{and} (\tau_{2})_{mn}$ are off diagonal matrices in the second line.And using the approximation for large volume
        ($v\gg 1$) we find
        
        \begin{eqnarray}
        \frac{2}{8\pi l_{p}^{2}\gamma\mu}\left(V_{v+\bar{\mu}}-V_{v-\bar{\mu}}\right)=sgn(\bar{p})V_{0}^{1/3}\sqrt{|\bar{p}|},
        \end{eqnarray}
        we conclude that
        
        \begin{equation}
                    C_{grav}^1=\frac{-3sgn(\bar{p})}{\kappa\gamma^{2}}\left[\left(\bar{k}\gamma V_{0}^{1/3}\right)^{2}-1/3\bar{\mu}^2\left(\bar{k}\gamma V_{0}^{1/3}\right)^{4}\right]V_{0}^{1/3}\sqrt{|\bar{p}|}
        \end{equation}
       \item For the second part of the constraint
     \begin{eqnarray} \label{93}
           {C}_{grav}^{2}&=&\frac{4isgn(\bar{p})}{\kappa^{2}\gamma^{3}{\bar{\mu}}^{3}}\sum_{ijk}\varepsilon^{ijk}tr\left(2\gamma^{2}{\bar{\mu}}^{2}V_{0}^{2/3}\tau_{i}\tau_{j}\hat{h_{k}}[\hat{h_{k}}^{-1},\hat{V}]\right)\\\nonumber&=& 24 \frac{isgn(\bar{p})V_{0}^3}{\kappa^{2}\gamma\bar{\mu}}tr\left[ \tau_{1}\tau_{2}{h_{3}}[{h_{3}}^{-1},{V}]\right]
    \end{eqnarray}.
    Using
    \begin{equation}
        \tau_{1}\tau_{2}=-\frac{1}{4}\begin{pmatrix} i&0\\0&-i
        \end{pmatrix}.
    \end{equation}
    
 we find 
\begin{equation}
tr\left[ \tau_{1}\tau_{2}\begin{pmatrix}V_{v}-V_{v+1}&0\\0&V_{v}-V_{v-1}\end{pmatrix}\right]=\frac{1}{4}i\left(V_{v+1}-V_{v-1}\right),
 \end{equation}
 and substituting in (\refeq{93})give us :
 \begin{equation}
     {C}_{grav}^{2}=-\frac{3V_{0}^{2/3}}{\kappa}\times2\frac{\left(V_{v+1}-V_{v-1}\right)}{8\pi \gamma l_{p}^2\bar{\mu}}=-\frac{3V_{0}}{\kappa}\sqrt{|\bar{p}|}
 \end{equation}
    \item  Finally, the effective Hamiltonian constraint is given by:
 \begin{eqnarray} \label{97}
     H^{eff}\nonumber&=&C_{grav}^{1}+C_{grav}^{2}+H^{m}\\\nonumber&=&-\frac{3V_{0}}{\kappa}\sqrt{|\bar{p}|} \left[1+
     \left(\bar{k}^{2}-1/3\bar{\mu}^2\bar{k}^4\gamma^2 V_{0}^{2/3}\right)\right]+H^{m}\\&=& -\frac{3V_{0}}{\kappa}\sqrt{|\bar{p}|}\left[1+\frac{sin^{2}(\gamma\bar{k}\bar{\mu} V_{0}^{1/3})}{\gamma^{2}\bar{\mu}^{2}V_{0}^{2/3}}\right]+H^{m},
 \end{eqnarray} 
 with $ \bar{\mu}=\frac{1}{V_{0}^{1/3}}\sqrt\frac{\Delta}{\bar{p}} $ and $\Delta$ presents the value of the minimum area in loop quantum gravity \cite{thiemann1998quantum}.
 
 and 
 \begin{equation}
     H^{m}= C^{m}=V_{0}\bar{p}^{3/2}\rho_{m}
 \end{equation}
 \end{enumerate}
 Now we can make a comparison between the last expression of the effective Hamiltonian constraint(97) with the classical hamiltonian  constraint (\refeq{eq(32)})in order to deduce the holonomy correction, which is precisely given by replacing $\bar{k}\to {sin(\gamma\bar{k}\Tilde{\bar{\mu}})}/{\gamma\Tilde{\bar{\mu}}}$ with $\Tilde{\bar{\mu}}=V_{0}^{1/3}\bar{\mu}=\sqrt{\Delta/|\bar{p}|}$.
 \subsection{Effective equation of motion}
 Now  we can use the effective Hamiltonian constraint(\refeq{97}) with the Poisson bracket Algebra(\refeq{reviewD}) to find the effective equation of motion.
 We have 
 \begin{equation}
     \dot{p}=\{p,H^{eff}\}=-\frac{2\sqrt{|\bar{p}}|}{\gamma\bar{\mu}V_{0}^{1/3}}sin(\gamma\bar{k}\bar{\mu} V_{0}^{1/3})cos(\gamma\bar{k}\bar{\mu} V_{0}^{1/3})
 \end{equation}
 and 
  \begin{equation}
  H^2=\frac{1}{4}\frac{\dot{\bar{p}}^2}{\bar{p}^2}=\frac{1}{{|\bar{p}}|\gamma^2\bar{\mu}^2 V_{0}^{2/3}}sin^2(\gamma\bar{k}\bar{\mu} V_{0}^{1/3})cos^2(\gamma\bar{k}\bar{\mu} V_{0}^{1/3})
  \end{equation}
 Using the total Hamiltonian density constraint 
   \begin{equation}
   H^{eff}=C^{T}=C_{grav}^{1}+C_{grav}^{2}+H^{m}=0.
   \end{equation}
 
 Which implies 
 \begin{equation}
     \frac{3V_{0}}{\kappa}\sqrt{|\bar{p}|}\left[1+\frac{sin^{2}(\gamma\bar{k}\mu V_{0}^{1/3})}{\gamma^{2}\bar{\mu}^{2}V_{0}^{2/3}}\right]=V_{0}\bar{p}^{3/2}\rho_{m}
 \end{equation}
 so that 
  \begin{equation}
\frac{sin^{2}(\gamma\bar{k}\mu V_{0}^{1/3})}{\gamma^{2}\bar{\mu}^{2}V_{0}^{2/3}}=\frac{1}{3}\rho_{m}\bar{p}\kappa-1
  \end{equation}
with $\bar{p}=a^{2}$. Substituting the last equation (103) in Eq.(100) we find

 \begin{eqnarray}
  H^2&\nonumber=&\frac{1}{{|\bar{p}}|\gamma^2\bar{\mu}^2 V_{0}^{2/3}}sin^2(\gamma\bar{k}\bar{\mu} V_{0}^{1/3})\left[1-sin^2(\gamma\bar{k}\bar{\mu} V_{0}^{1/3})\right]\\&=&\left(\frac{1}{3}\rho_{m}\kappa-\frac{1}{a^2}\right)\left(1+\bar{\mu}^2V_{0}^{2/3}\gamma^2-\frac{1}{3}\rho_{m}\kappa\bar{p}\bar{\mu}^2V_{0}^{2/3}\gamma^2\right),
 \end{eqnarray}
  which give us the modified Friedmann equation.
  Now if we use $\bar{\mu}^2=\frac{l_{min}^2}{V_0^{2/3}|\bar{p}|}$
    we can rewrite the modified Friedmann equation (104)in the following form: 

 \begin{eqnarray} \label{105}
    H^2&\nonumber=&\left(\frac{1}{3}\rho_{m}\kappa-\frac{1}{a^2}\right)\left(1+\frac{l_{min}^2\gamma^2}{a^{2}}-\frac{\kappa l_{min}^2\gamma^2} {3}\rho_{m}\right)\\&=&\frac{\kappa}{3}\left(\rho_{m}-\frac{3}{\kappa a^2}\right)\frac{\kappa l_{min}^2\gamma^2}{3}\left(\frac{3}{\kappa l_{min}^2\gamma^2}+ \frac{3}{\kappa a^2}-      \rho_{m} \right)
  \end{eqnarray}
  where 
  \begin{equation}
      \rho_{cl}=\frac{3}{\kappa a^2}
  \end{equation}
 
 is the classical solution where the scale factor reaches its maximum value,the matter density reaches its minimum value, and the classical dynamics exhibits a turning point from an expanding phase to a contracting phase.
 And 
 
 \begin{equation}
      \rho_{crit}=\frac{3}{\kappa l_{min}^2\gamma^2}=0.82 \rho_{pl}
  \end{equation}
  Our study show that quantum dynamics returns this point but also gives rise to additional solutions resolving the (Big-Bung) and the big crunch singularities.
  The second quantum critical density is given by
  
  \begin{equation}
      \rho_{c}=\frac{3}{\kappa l_{min}^2\gamma^2}\left(1+\frac{l_{min}^2\gamma^2}{a^2} \right)
  \end{equation}
   Similarly we can use the Poisson bracket(\refeq{reviewD}) to determine $\dot{v}$

  \begin{eqnarray}
      \dot{v}=\{v,H^{eff}\}&=&\nonumber\frac{\kappa}{3V_0}\frac{\partial H^{eff}}{\partial \bar{k}}\frac{\partial v}{\partial \bar{p}}\\ &=& 3\frac{|\bar{p}|}{V_0\bar{\mu}}sin(\gamma\bar{k}\bar{\mu} V_{0}^{1/3})cos(\gamma\bar{k}\bar{\mu} V_{0}^{1/3}),
  \end{eqnarray}
  we can also use the equation (\refeq{44}) to determine the specific relation
    \begin{equation}
         \frac{dv}{d\Phi} =3\frac{|\bar{p}|}{V_0\bar{\mu}}sin(\gamma\bar{k}\bar{\mu} V_{0}^{1/3})cos(\gamma\bar{k}\bar{\mu} V_{0}^{1/3})\frac{\bar{p}^{3/2}}{\bar{\Pi}_\phi}=\frac{\bar{p}^{3/2}}{\bar{\Pi}_\phi}\dot{v}=\frac{\bar{p}^{3/2}}{\bar{\Pi}_\phi}3Hv
    \end{equation}
  
 where we have used
  \begin{equation}
    \frac{\dot{v}}{v}=3H
   \end{equation}
 
 Using the modified Friedmann equation (\refeq{105}) straightforward calculus gives
 
 \begin{equation}
 \frac{dv}{d\Phi}=\pm\sqrt{12\pi G}\frac{1}{\sqrt{ \rho_{crit}}}\left( 1-\frac{6v^{4/3}}{{\kappa \bar{\Pi}_{\phi}}^2}\right)^{1/2}\left( \rho_{crit}v^{2}+\frac{3}{\kappa}v^{4/3}-\frac{\bar{\Pi}_{\phi}^2}{2}\right)^{1/2}
   \end{equation}
  
  From the first root of the effective Friedmann equation (112) we find
  \begin{equation}
      v_{max}=\left(\frac{8\pi G}{6} \right)^{3/4}\bar{\Pi}_{\phi}^{3/2}
  \end{equation}
 Thus for k=+1 we have a collapse at $v=v_{max}$.\\
  From the second root of the effective Friedmann equation we find 
 
 \begin{equation}
      v_{min}=\frac{\bar{\Pi}_{\phi}}{\sqrt{2.\rho_{crit}}},
    \end{equation}
    which is the minimum value for the volume at the bounce.
  As another illustration, we consider a semi classical state representing a large universe at the classical collapse with  a density about the current density of our universe $\rho_{min}=9.7.10^{-30}gm/cm^{3}=15.10^{-124}\rho_{p}$.In order to calculate $v_{min}$in this case at the bounce , we need first to calculate $\bar{\Pi}_{\phi} $.Using the equation(\refeq{42}) at  the classical collapse and the Eq.(113) of  $v_{max}$ we find

\begin{equation}
\rho_{min}=\frac{\bar{\Pi}^{2}_{\phi}}{2v_{max}^{2}}\implies \bar{\Pi}_{\phi}=\frac{1}{2\rho_{min}\left(\frac{8\pi G}{6}\right)^{3/2}}=3,9.10^{121}l_p
    \end{equation}
Now using the value of $\bar{\Pi}_{\phi}$ in Eq.(115) and the Eq.(114) we find the value of $v_{min}$ at the bounce

\begin{equation}
v_{min}=0,78l_{p}^2\bar{\Pi}_{\phi}=3.10^{121}l_p^{3}=1,2.10^{23}cm^{3}
    \end{equation}

     \begin{figure}[h] 
    \centering
        \includegraphics [scale=0.40]{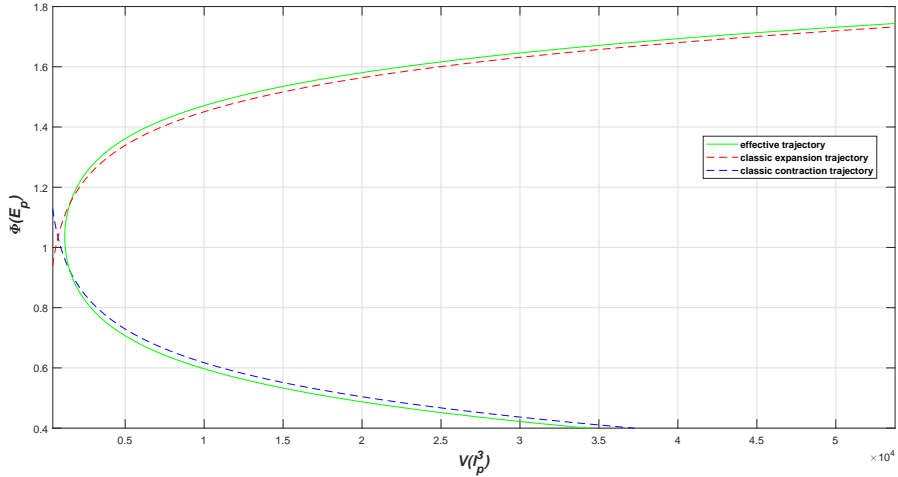} 
    \caption{A comparison of the effective dynamic with the classic dynamics (dashed curve) is presented near the bounce point.The classical trajectories deviates significantly from the effective evolution near the bounce and evolves into singularities.In this figure we have assumed $\bar{\Pi}_{\phi}=1500l_p $ and $\rho_{crit}=0.82 \rho_{pl}$}
    \label{fig:my_label2}
     \end{figure}

         \begin{figure}[h]
     \centering
    \includegraphics [scale=0.40]{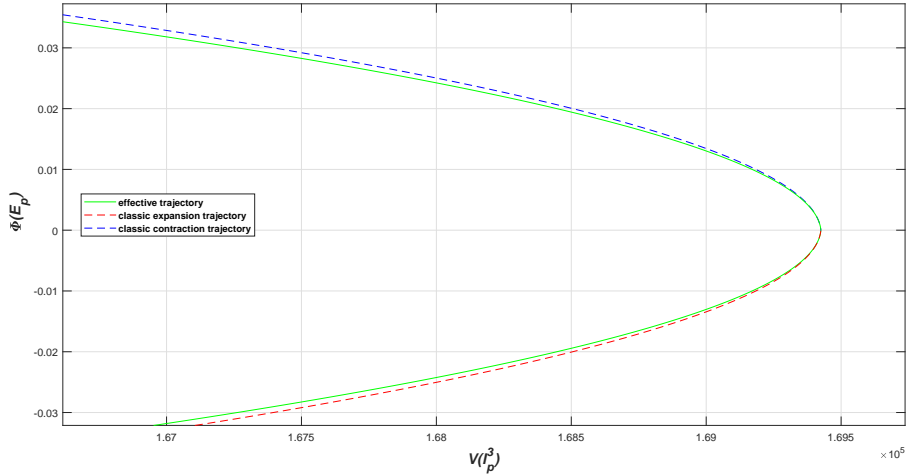} 
    \caption{A comparison of the effective dynamic with the classic dynamics (dashed curves) at large volume and  near the  collapse point, when space-time curvature is significantly weaker then the Planck scale .In this figure we have assumed $\bar{\Pi}_{\phi}=1500l_p $ and $\rho_{crit}=0.82 \rho_{pl}$}
    \label{fig:my_label}
       \end{figure}

 \section{Discussion} \label{sectionC}
 
We have constructed the effective (semi-classical) equation of motion of the closed (k=+1) F.R.W model using the framework of Loop quantum cosmology .In particular we have used the K-Quantization method and the improved dynamic $\bar{\mu}-{sheem}$.
Choosing the K-quantization method which depends on the holonomy of the extrinsic curvature as basic variables instead of the holonomy of the connection used in the literature \cite{ashtekar2007loop} somewhat facilitates the quantization process  as we don't need to use the left and right invariant (vector-fields) to define a Holonomy-operator. It should be noted that the difference between the two methods in the case of a closed metric (k=+1)leads to a different effective equation of motion as well as the results deducted from it.\\
The key features of the K-quantization method applied to the closed model can be summarized as follows:\\
After we had derived the effective Hamiltonian constraint with a massless scalar field and the holonomy correction expression, we found the effective equation of motion.In this paper we have used a massless scalar field,so that we are ignoring the inverse volume corrections to the matter Hamiltonian.In the presence of potentials,the result obtained will significantly change.note  also that this scalar field $\phi$ serves as emergent time in both classical general relativity or in the effective theory, in particular to deduce the effective equation of motion.

With the effective Hamiltonian, we have also derived an effective Friedmann equation which altered the classical dynamic and gave us  a first root responsible for the recollapse where the minimum density is given by $ \rho_{min}={\bar{\Pi}^{2}_{\phi}}/{2v_{max}^{2}}$ and       $v_{max}= \left({8\pi G})\right/6)^{3/4}\bar{\Pi}^{3/2}_{\phi}$.The effective trajectory is highly peaked in a classical trajectory in the low energy regime, when space-time curvature is significantly weaker then the Planck scale so that we have a viable IR(infrared) behavior where general relativity provide a good  approximation  see {\textbf{figure (\ref{fig:my_label}})}. In addition, there is a second root responsible for the bounce;where the effective equation predicted that that during the contraction of the universe with a massless scalar field the energy increases until it reaches its maximum density at the bounce  $ \rho_{max}={\bar{\Pi}^{2}_{\phi}}/{2v_{min}^{2}}$ with a minimum volume $ v_{min}={\bar{\Pi}_{\phi}}/{\sqrt{2.0,82\rho_{p}}}$ see {\textbf{figure (\ref{fig:my_label2}})}( cure the UV difficulties by resolving the singularity).\\

  In the effective Hamiltonian we have derived an effective Friedmann equation, which gave us constant critical density $ \rho_{crit}=0.82 \rho_{pl}$. 
  We should remember that in order to obtain the semi classical limit of the hamiltonian constraint we have constrained the value $\Bar{\mu}-{scheme}$  using the minimal area $ \Delta=A_{min}=2\pi\sqrt{3}\gamma l_{p}^{2} $ and we have also used the condition $\bar p=a^2\gg 1 $.
  In {\textbf{figure (\ref{fig:my_label2}})} we notice that the value of the volume at the bounce is equal to $v_{min} =1171 l_p^3 $and this value is sufficiently greater than 1, where we have assumed $\bar{\Pi}_{\phi}=1500l_p $.
   In order to obtain practical values we can use the current minimum density $\rho_{min}=9.7.10^{-30}gm/cm^{3}$ in closed metric therefore the minimum value for the volume at the bounce is equal to  $v_{min}=1,2.10^{23}cm^{3}$ with $a_{min}=1,2.10^{15}cm^{3}\gg 1$,and this means that the condition that we have used to constrain the value of $\overline{\mu}$ in the semi-classical limit is consistent with the numerical results.

  The effective continuous picture described in this paper could possibly extended to a perturbation setting.For instance we can use the correction of holonomy  in the perturbation setting by  replacing $\bar{k}\to {sin(\gamma\bar{k}\Tilde{\bar{\mu}})}/{\gamma\Tilde{\bar{\mu}}}$ (Eq.(97) with $\Tilde{\bar{\mu}}=V_{0}^{1/3}\bar{\mu}=\sqrt{\Delta/|\bar{p}|}$) as in the case of  scalar or vector or tensor perturbations in the flat space .In principle one could then calculate modifications to the classical perturbation equations and examine their consequences . Understanding perturbative theory with LQC would be essential to test the model of LQC and also to understand the past of our universe.In fact the theory would have to predict a nearly scale invariant spectrum of primordial fluctuation. This is a topic for a future research in k=+1 RW cosmology.

\end{document}